\documentclass{article}
\usepackage{amssymb}
\usepackage{epsf}
\begin{document}
\baselineskip7mm
\title{New properties of scalar field dynamics in brane isotropic
cosmological models}
\author{A.V. Toporensky, P.V. Tretyakov and V.O. Ustiansky}
\date{}
\maketitle
\hspace{8mm}{\em Sternberg Astronomical Institute,
 Moscow 119899, Russia}
\begin{abstract}
Several aspects of scalar field dynamics on a brane which differs
from corresponding regimes in the standard cosmology are investigated.
We consider asymptotic solution near a singularity, condition for
inflation and bounces and some detail of chaotic behavior in the
brane model. Each results are compared with those known in the
standard cosmology.
\end{abstract}

\section{Introduction}

The idea that we live on a three-dimensional brane embedded in a
higher-dimensional space-time have attracted a lot of interests for the past few
years. This idea has a long history (for references see~\cite{Visser:qm}) but
its recent renewal caused primarily by the the works of L.\,Randal and
R.\,Sundrum~\cite{Randall:1999ee,Randall:1999vf}.

In such brane-world scenarios the ordinary matter is confined to the brane
while the graviton can propagate in a whole spacetime. The effective
four-dimensional gravity on the brane is thus modified by extra
terms~\cite{Binetruy:2000hy,Shiromizu:1999wj},
which may play an important role in cosmology. One of the predictions of
brane-world models is that the contribution from the energy-momentum
tensor components in the equation of
motion contains both linear and second order terms while in conventional
cosmology they contribute only linearly.  The other prediction is the
presence of non-local terms that are the projection of a Weyl tensor in the
bulk onto the brane.  In the case of FRW metrics on a brane
the bulk contributes the same way as a radiation. That is why
this term is called ``dark radiation'',  despite its pure geometrical
nature.

Dynamics of a brane Universe filled with a perfect fluid have been
intensively investigated during last three years (see, for example,
\cite{Maartens, C-S1, C-S2})
and some new regimes that are not inherent to the standard cosmology, such as
stable oscillation \cite{C-S2}, collaps of a flat Universe \cite{Santos},
stage of growing anisotropy in Bianchi I Universe \cite{Topor},
unusual analog of an Einstein static Universe \cite{Ge} were found.

A scalar field dynamics is complicated even in the standard cosmology
\cite{Star},
so, it is not surprizing that the case of scalar field on a brane gives us
a variety of very interesting dynamical regimes. Some features of a brane
inflation
have been studied in \cite{Maartens:1999hf, Varun, Copeland:2000hn},
the cosmological dynamics for exponential scalar field
potentials have been described in
\cite{Maeda,Dunsby}.  Currently the problem of a scalar field dynamics on a brane
continues to attract a great attention.

In the present paper we describe how several known regimes of the standard
scalar field cosmology are modified in the brane scenario.
We consider the effective four-dimensional FRW cosmology on the
brane with a scalar field $\varphi$ with potential $V(\varphi)$.
In Sec.2 modifications mostly arising from the square energy-momentum
corrections to the Einstein equations on a brane are described. In Sec.3
we describe some dynamical features arising from the presence of
a negative ``dark radiation'' term.

\section{High-energy brane dynamics}

The effective four-dimensional equations of motion
(this form of equations can be easily derived from those in
\cite{Binetruy:2000hy}) are
\begin{eqnarray}
\frac{\dot a^2}{a^2}+\frac{k}{a^2}&=&
\frac{\tilde \kappa^2}{3}\left(\frac12\dot\varphi^2+V\right)
+\frac{\kappa^4}{36}\left(
\frac12\dot\varphi^2+V\right)^2+\frac{C}{a^4}\label{constr}\\
\frac{\ddot a}{a}+\frac{2\dot a^2}{a^2} + \frac{2k}{a^2}&=&
\tilde \kappa^2 V+\frac{\kappa^4}{48}(4V^2-\dot \varphi^4)
+\frac{C}{a^4}\\
\ddot\varphi&=&-3\frac{\dot a}{a}\dot\varphi-V',\label{ddotphi}
\end{eqnarray}
where dot (prime) denotes differentiation with respect to time ($\varphi$),
$\tilde \kappa^2=8\pi/M_{(4)}^2$, $\kappa^2=8\pi/M_{(5)}^3$. Here $M_{(5)}$
is the fundamental 5-dimensional Planck mass, $M_{(4)}$ is an effective
4-dimensional Planck mass on a brane which depends on a brane tension
$\lambda$ as \cite{Maartens}
\begin{equation}
M_{(4)}=\sqrt{\frac{3}{4\pi}}\left(\frac{M_{(5)}^2}{\sqrt{\lambda}}\right)
M_{(5)}.
\end{equation}

 The brane tension
$\lambda$ is restricted to be positive in order to have a correct sign of
gravitational force. The ``dark radiation'' constant $C$ has no restrictions
on its sign.

One of the distinctive features of equation (\ref{constr}) is the presence
of term quadratic in the field energy density.  The contribution of this
term should be negligible at low energies in order to reveal standard
cosmology~\cite{Cline:1999ts,Binetruy:2000hy}, but at high energies this
term becomes dominative.  So let us investigate the set of equations
(\ref{constr})--(\ref{ddotphi}) in the high-energy limit. We remind a reader
that a scalar field dynamics is quite different in the stage of contracting
and expanding of the Universe. In the standard cosmology
(we use the common units $M_{(4)}/\sqrt{8 \pi}=1$ in the standard case
and keep all dimension parameters explicitly written in the nonstandard one)
for the contraction stage
if the potential $V(\varphi)$ is less steeper than $V(\varphi) \sim \exp{
\sqrt{6} \varphi}$
then we typically have the same regime as for $V(\varphi)=0$
with
$a(t) \sim t^{1/3},$
and a scalar field diverging as $\varphi(t) \sim t^{-1}$ \cite{Star}.
For steeper potential it becomes dynamically important and
the scalar field diverges in an oscillatory regime 
 \cite{Foster}.

The picture changes dramatically in the brane scenario. Indeed, for
$V(\varphi)=0$
  substituting the
ansatz $$ a=At^\alpha, \qquad\varphi=Bt^\beta+\varphi_0 $$
into the equations of motion  one may obtain the
following solution:
\begin{equation}
a(t)=At^{1/6},\qquad\varphi(t)=\pm\frac{M_{(5)}^{3/2}}{\sqrt{2\pi}}
t^{1/2}+\varphi_0,
\qquad t\to0,\label{asympt}
\end{equation}
where $A>0$ and $\varphi_0$ are the constants of integration.  One can see
that the principal
feature of the model under consideration is that the scalar field tends to a
constant value while in conventional cosmology it tends to infinity.
Although $\varphi$ is finite, its time derivative tends to infinity providing
``stiff fluid'' equation of state $p=\varepsilon$ as for a massless scalar
field.  From the other side, as $\varphi$ remians finite while $\dot
\varphi$ diverges, then the kinetic energy of the scalar field always
dominates the potential energy near a singularity at the contraction stage
and the asymptotic (5) is valid for an arbitrary $V(\varphi)$.

At the expansion stage a typical regime for a scalar field is inflation.
Let us now consider the problem of the
steepness of the scalar field potential $V(\varphi)$ allowing the existence of
inflationary stage in the evolution of a brane world.  On an inflationary stage
we can neglect contributions from the spatial curvature and the dark radiation terms in
the equations of motion.

The equations for two slow-roll parameters
can be written in the following
form (see the high-energy limit of equations in \cite{Maartens:1999hf})
\begin{equation}
\epsilon\simeq
\frac{3 M_{(5)}^6}{16 V \pi^2}\left(\frac{V'}V\right)^2
\end{equation}
and
\begin{equation}
\eta\simeq \frac{3 M_{(5)}^6}{16 V \pi^2}\left(\frac{V''}{V}\right)
\end{equation}
where $\simeq$ denotes equality within the slow-roll approximation.
The condition for inflation is $|max(\epsilon, \eta)| < 1$.
In the standard cosmology this condition is
violated for potentials steeper than $V(\varphi)\sim
e^{\sqrt2\varphi}$.  In our case this condition
can be violated only for potentials
steeper than
$
V(\varphi)\sim(\varphi-\varphi_0)^{-2}\
$
and for the critical case of
$
V(\varphi)=A/(\varphi-\varphi_0)^{2}
$
the condition $|max(\epsilon, \eta)| < 1$ is violated for
 $A<9 M_{(5)}^6/(8 \pi^2)$.

So in the brane cosmology inflation requires less restrictions on the
shapes of potential than in the conventional case, admitting
potentials in the form of a ``potential wall''.

 The nonstandard dependence on the components of the energy-momentum tensor
of the matter alters also the conditions for a chaotic regime in this model.
In conventional cosmology it has been shown that for a wide class of
potentials $V(\varphi)$ the dynamics of a closed model ($k=+1$) is
chaotic~\cite{Cornish:1997ah,Kamenshchik:1998ue}.  Without loss of generality
we can fix the initial value of $\dot a$ to be zero. The initial value of
$\dot \varphi$ can be determined from the equation (\ref{constr}) and
thus we have two-dimensional initial condition space. The numerical
calculations show that starting from some initial conditions the phase trajectory
can either fall into singularity or can pass through a local minimum of the
scale factor, which is often referenced to as a ``bounce''.  In the latter
case the scale factor will inevitably have a local maximum and thus we
returns to the picture similar to the initial one with two possible outcomes.
It has been shown that there is a set of infinitely bouncing solutions those
initial condition measure is equal to zero.

On a $(a,\varphi)$ plane the region where the scale factor can have extremum
is determined from the constraint (1) by the inequality
\begin{equation}
V(\varphi) \le \frac{3 M_{(5)}^3}{4 \pi a^2} \left(1-\frac{C}{a^2}\right). \label{evklid}
\end{equation}

 We remined
a reader that in standard cosmology chaos may disappear for exponential and
steeper than exponential potential of a scalar field. In Ref. \cite{topor}
an analytical approach to this problem has been developped. It has been shown
that using two facts established in numerical experiments, namely
\begin{description}
\item[(a)] all simple periodical trajectories have a point with $\dot \varphi
=0, \dot a=0$ at the boundary, corresponding to equality sign in (8),
\item[(b)] all trajectories which have in this reversal point a velocity
vector directing into the region (8) fall into a singularity and, then,
can not be periodical trajectories,
\end{description}
it is possible to write a rather simple necessary condition for existence
of chaos.

These two facts may be violated in so called strong chaotic regime, however,
this regime exists only for potential at least less steep than a quadratic
one \cite{Pavluchenko:1999zv}. So, we can use them for steep potentials when the chaotic
regime begin to disappear.

Now we apply the technique of Ref. \cite{topor} to a brane case.
First we assume that $C=0$.
A trajectory
with $\dot \varphi=0$ and $\dot a=0$ is directed outside of region (8)
if
\begin{equation}
|\frac{\ddot a}{\ddot \varphi}| > |\frac{da(\varphi)}{d\varphi}|
\end{equation}
where $a(\varphi)$ is the equation of the boundary. Substituting
$\ddot a$ and $\ddot \varphi$ from Eqs.(2)-(3) and using Eq. (1) we get
this condition in the form
\begin{equation}
\frac{\kappa^4}{36} V(\varphi)^3 > \left(\frac{dV(\varphi)}{d\varphi}\right)^2.
\end{equation}

If this condition is violated for all $\varphi$, chaos definitely
disappears.
As the power index of potential $V(\varphi)$ in (10) is bigger than the
power index of the derivative $V'(\varphi)$, the inequality (10) is
satisfied for exponential potentials if we take a suffitiently large
$\varphi$. Moreover, it remaines true for potentials like
$V \sim \exp(\varphi^2)$. As a result, we can see that a chaotic dynamics
on a brane can take place for steeper potential than in standard
cosmology.

Potentials, violating (10) must have the form of potential wall. A critical
case has the form
\begin{equation}
V(\varphi)=\frac{A}{(\varphi-\varphi_0)^2}.
\end{equation}
It can be easily seen by integrating of (10) that for $A<9 M_{(5)}^6/(4 \pi^2)$ this
condition is not satisfied for all $\varphi$. Our numerical investigations
indicates that chaos is absent in this case.

The same analysis can be carried out for a nonzero $C$
though the corresponding equations  become less simple in this case. A general property
is that positive $C$ restricts condition for the chaotic behavior and
negative $C$ makes it weaker. In the limit $C \to -\infty$ we can wright
the simple analog of (10)
\begin{equation}
\frac{\kappa^4}{9} V(\varphi)^3 > \left(\frac{dV(\varphi)}{d\varphi}\right)^2
\end{equation}
which lead to a critical potential of the form (11) with $A=9M_{(5)}^6/(16
\pi^2)$.

We can summarise these results in the following picture. It is known that
in the standard scalar field cosmology there is a substantial differense
between potentials less steep  and steeper than the exponential one.
For steep enough potentials the inflation become impossible, the chaotic regime
for positive spatial curvature disappears and the behavior of a scalar field
during a contraction of the Universe becomes infinitely oscillating. For the
latter modification we have no analog in the brane case - a scalar field
do not diverges at the contraction stage independently on the potential.
The first two modifications of the scalar field dynamics for steep potentials,
however, have their analogs in the brane scenario. The boundary case which
separates two different cases of the dynamics (analogous to exponential
potentials in the standard cosmology) is a ``potential wall'' of the form
$V(\varphi) \sim 1/(\varphi-\varphi_0)^2$.

\section{Effects of a negative ``dark radiation''}

Now let us consider the effects originated from the presence of a ``dark
radiation'' term. The important feature of the ``dark radiation'' term is that
the the constant $C$ can be either positive or negative.
Due to the negative $C$ the evolution of the
scale factor can be non-monotonic not only in the positive curvature case but
also in flat and negative curvature case.
Indeed, as we can see from the system (\ref{constr})--(\ref{ddotphi}) the $C$-contaning terms can
mimic curvature terms. This leads, in particular, to existence of
bounces in flat Universe. Moreover,  probability of a bounce increases
with increasing of $|C|$ and can reach  considerable values.

To illustrate this point we have performed numerical integration of system
(\ref{constr})--(\ref{ddotphi}) in a pure nonstandard regime ($\rho \gg \lambda$)
starting from some initial scale factor at the stage of contracting.
In our numerical calculations we put $\kappa =1$.
A particular trajectory is determined by two values - initial scalar field
and its derivative. After fixing initial $\varphi$ and $\dot \varphi$ the
initial $\dot a$ can be found from the constraint equation (\ref{constr}).
 We have studied
a compact area in the initial condition plane $(\varphi, \dot \varphi)$
which is allowed by the constraint (1) and satisfying the inequality
$$
\frac{m^2 \varphi^2}{2} + \frac{\dot \varphi^2}{2} < 1.
$$
Introducing the homogenious measure on this compact set of initial condition,
we have the following results. For the first set of
the bounce probabilities $P$ the mass of the scalar field is
chosen to be $0.1$.  As for the flat
Universe $C$ and $a$ enter in the equation of motion only in the combination
$C/a^4$, our results are independent on a particular initial
scale factor.
\hspace{0.6cm}
 \begin{center}
 \begin{tabular}{|c|c|c|c|c|c|}
 \hline
$C/a^4$& $-10^{-5}$  & $-10^{-4.5}$  & $-10^{-4}$ &  $-10^{-3}$ & $-10^{-2}$\\
 \hline
 P&$7.4\cdot10^{-4}$&$2.0\cdot10^{-3}$&$5.6
\cdot10^{-3}$&$9.2\cdot10^{-2}$&
$3.4\cdot10^{-1}$\\
 \hline

 \end{tabular}
 \end{center}

The dependence of the bounce probability on the scalar field mass $m$ is rather
complicated, though for small enough $m$ this dependence disappears. Some
values of $P$ for different $C$ and small $m$ ($m$ is less than $10^{-3}$)
in this situation are presented below.
\begin{center}
              \begin{tabular}{|c|c|c|c|}

              \hline

              $C/a^4$&$-10^{-4}$&$-10^{-3}$&$-10^{-2}$\\

              \hline

              P&0.029&0.072&0.195\\

              \hline

              \end{tabular}
              \end{center}

This regime of transition from contraction to expansion of the flat
Universe can exist only with some kind of negative energy terms in (\ref{constr}).
Brane cosmology gives us a simple example of such kind of energy sourses.
Our numerical studies show
that the measure of initial condition leading to a bounce can be
significant for large negative $C$.

\begin{figure}[hb]
\epsfxsize=0.5\hsize
\centerline{\epsfbox{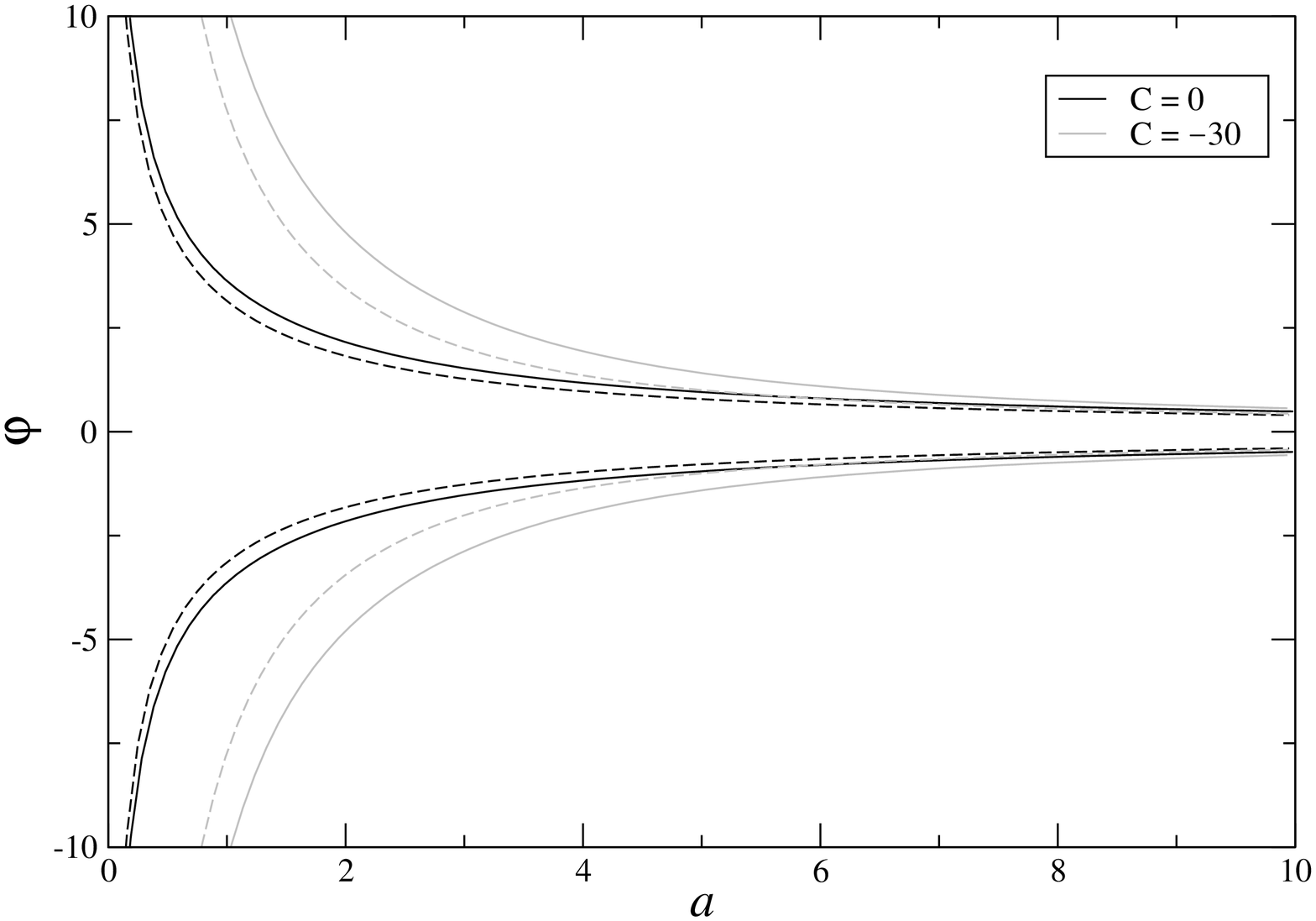}}
\caption{The regions where the scale factor of closed universe can have extremum
are bounded by a solid lines. The regions where the scale factor can have local
minimum are bounded by solid and dashed lines. Black color corresponds to
$C=0$ and gray color corresponds to $C=-30$\@.}
\label{boundaries}
\end{figure}

The dynamics of a brane universe in positive curvature case is more complicated.
A positive $C$ leads to shrinking the
regions of possible extrema of the scale factor
 and for large enough $C$ the chaos disappears completely.
The dynamical picture is the same as in the standard case of a scalar field
dynamics in the presence of a hydrodynamical matter with the equation
of state $p=\epsilon/3$ \cite{entropy}.
A negative $C$ (in this case we have no direct analogy with
a convensional cosmology)
acts in opposite direction:  the region of possible extremum of the scale
factor grows with the growth of the absolute value of $C$ and (as we will see
later) the chaotic properties of the universe increase. In 
Fig.~\ref{boundaries} the region where the scale factor can have extremum
is bounded by two solid lines and the region where the scale factor
can have minimum is bounded by dashed and solid lines. For this figure
the scalar field potential is chosen to be $V(\varphi)=\frac{m^2\varphi^2}2$
where $m=0.5$ and the value of brane tension $\lambda=1$
(we return to the units $M_{(4)}/\sqrt{8 \pi}=1$). The black color
corresponds to $C=0$ while the gray color corresponds to $C=-30$.

\bigskip

We describe below an interesting transformation of the chaotic
dynamics that takes place in the standard regime ($\rho \ll \lambda$).
From now on let us assume the scalar field potential to be in a form
$V(\varphi)=\frac{m^2\varphi^2}2$\@. In
the case $C=0$ the initial conditions that lead to bounce forms a set of a
narrow regions separated by
a wide regions where solution is singular. This picture undergoes 
substantial transformations with growing $|C|$.
 In Fig.2 we fixed
the initial value of $\varphi$ to be zero and we choose $m=0.5$,
 while allowing $C$ and $a$ to change in the interval from
$-100$ to $0$\@ and from $1$ to $30$ respectively. The initial conditions
that leads to bounce after start from $\dot a=0$ 
are colored in black while the white color corresponds
to singular solutions.  One can see that with the growth of the absolute
value of $C$ the intervals of the initial scale factor
 that lead to bounce become wider, and for
$C\approx-15$ the first two intervals merge. 
 Solutions that start in the region of merging become
extreemly chaotic. Such solutions can oscillate for a long period
without leaving a high-curvature region.
 With further decreasing of $C$ one
can see successive interval merging and monotonic growth of the measure
of initial conditions that lead to bounce.

  This picture is similar to one
previously found in standard cosmological models with gently sloping
potential~\cite{Pavluchenko:1999zv} and in models that include second order
curvature corrections~\cite{Alekseev:eh}. In brane scenario this effect
exists for a simple $m^2 \varphi^2/2$ potential. It is interesting that
for a pure nonstandard regime $(\rho \gg \lambda)$ this strong chaos regime
exists only for potentials significantly less steep than a quadratic one.

\begin{figure}[hb]
\epsfxsize=\hsize
\centerline{\epsfbox{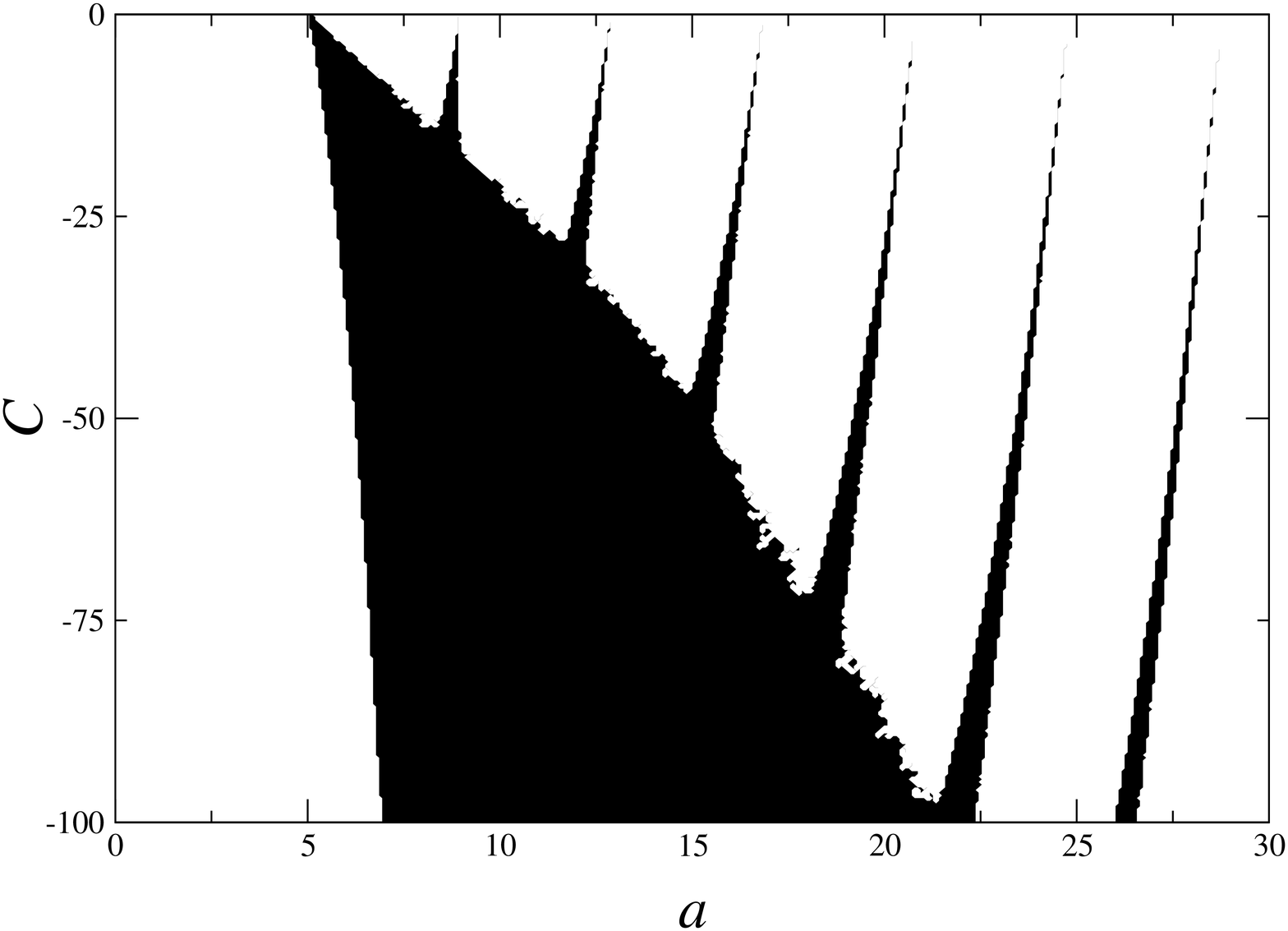}}
\caption{The $\varphi=0$ crossection of initial conditions space for different
values of $C$ that lead to bounce.}
\label{intervals}
\end{figure}

\section*{Acknowledgments}

The work was partially supported by RFBR grants Ns. 00-15-96699
and 02-02-16817.

\end{document}